\documentclass[rmp,nofootinbib,amsmath,amssymb,twocolumn]{revtex4}
\usepackage{graphicx} 

\begin{document}
\bibliographystyle{apsrev4-1}

\setcitestyle{numbers,square,citesep={,}}
\title{\textbf{Efficient configurational-bias Monte-Carlo simulations of chain molecules with `swarms' of trial configurations}}
\author{Niels Boon}
\affiliation{Division of Physical Chemistry, Department of Chemistry, Lund University, SE-22100 Lund, Sweden}
\begin{abstract}
Proposed here is a dynamic Monte-Carlo algorithm that is efficient in simulating dense systems of long flexible chain molecules. It expands on the configurational-bias Monte-Carlo method through the simultaneous generation of a large set of trial configurations. This process is directed by attempting to terminate unfinished chains with a low statistical weight, and replacing these chains with clones (enrichments) of stronger chains. The efficiency of the resulting method is explored by simulating dense polymer brushes. A gain in efficiency of at least three orders of magnitude is observed with respect to the configurational-bias approach, and almost one order of magnitude with respect to recoil-growth Monte-Carlo. Furthermore, the inclusion of `waste recycling' is observed to be a powerful method for extracting meaningful statistics from the discarded configurations. 

\end{abstract}
\maketitle
\section{Introduction}
Polymer chains are challenging to model with computer simulations due to the vast number of possible chain configurations that quickly increases with the number of monomers. Additionally, the exploration of phase space is hindered by chain entanglements. While Molecular-Dynamics (MD) methods can be applied to follow the natural time evolution of such systems, Monte-Carlo(MC) approaches enable the introduction of unphysical `moves' between configurations, which may increase the rate of generating uncorrelated configurations\cite{Grest1995,Newman1999}. The MC method comprises two categories of approaches to sampling phase space. A \emph{static} MC algorithm explores phase space through successive generation of uncorrelated configurations from scratch\cite{Frenkel2002,Frenkel2006}. \emph{Dynamic} MC algorithms, on the other hand, will attempt to generate a new configuration based on the existing state. This characterization refers to the creation of a Markov chain of configurations \cite{Metropolis1953} in contrast to the physical dynamics of the system. \\

The `simple sampling' of polymer configurations by the consecutive adding of monomers at random orientations to form a chain is an example of a static approach. Simple sampling is, however, not efficient because most generated configurations will have a vanishing Boltzmann weight due to overlaps between monomers. This obstacle can be partially avoided by considering multiple `probe' positions for every monomer that is added to the growing chain, and select from those with a probability proportional to their (resulting) Boltzmann weight. That approach, which largely avoids monomer positions that lead to overlaps, is known as the Rosenbluth-Rosenbluth (RR) method. It requires keeping track of a statistical weight $W_i$ to remove the introduced biasing in this selection process of a polymer-chain configuration. Equilibrium properties $\langle A \rangle$ follow from summing over all generated configurations $i$
\begin{equation}
\langle A \rangle = \frac{\sum_i  W_i A }{\sum_i  W_i }. \label{eq:av}
\end{equation}
It was found, however, that for longer chains the RR method yields a very wide spread in the weights $W_i$, such that only a few configurations dominate the weighed average in Equation~(\ref{eq:av}). The simulation, therefore, will spend most of its time on configurations that do not contribute significantly to this weighted average\cite{Batoulis1987}. The pruned-enriched Rosenbluth method (PERM) was tailored to address this inefficiency\cite{Grassberger1997}. This method controls the generation (growth) of the chain configurations by attempting to terminate(prune) unfinished chains with a below-average weight. On the other hand, those with an above-average weight are cloned(enriched) such that two copies of the incomplete chain continue to grow with half of the original weight each\cite{Wall1959}. Pruning and enrichment maintains the correct statistics while more computational time is spent on polymer configurations of significant weight. This leads to a much more efficient sampling of phase space. The PERM approach enabled the sampling of very long chain molecules\cite{Grassberger1997, Frauenkron1997,Barkema1998} and protein structures using minimalist models\cite{Bastolla1998}.\\

Static algorithms such as RR and its extension PERM are not favorable for systems with multiple polymers as they involve finding a new configuration from scratch for \emph{all} chains in the system at the same time. This approach quickly becomes ineffective as the number of chains in the system is increased beyond one. It is, however, possible to render the RR method dynamic and the resulting approach is known as the configurational-bias Monte-Carlo (CBMC) method \cite{Siepmann1992,Depablo1992,Vlugt1998, Frenkel1999a,Frenkel2002}. Each step in the CBMC algorithm applies the RR method to generate a new configuration for a randomly chosen chain in the system. The existing configuration is then also `retraced' (re-weighed) such that the acceptance probability of the new configuration can be determined by comparing the weights of the new and the existing configuration. The CBMC method has proven to be very effective for finding properties of chain molecules such as alkanes\cite{Smit1995,Martin1998,Vlugt1999,Wu2012,Dubbeldam2013,Krishna2013,Sepehri2014}, and has been extended to wide array of other systems\cite{Dellago1998,Bolhuis1994,Biben1996,Dijkstra1994,Shelley1995,Shelley1994}.\\


This work aims at incorporating pruning and enrichment into the CBMC method. The approach is, nevertheless, quite different from earlier attempts such as DPERM\cite{COMBE2003}, for which only a marginal increase in efficiency w.r.t. CBMC was reported. An essential element is the \emph{simultaneous} generation (growth) of a fixed number of `candidate' configurations, while at the same time the reference configuration is retraced. This synchronized growth (and retracing) of candidate chains enables comparison of the weights of unfinished configurations. Pruning or enrichment can, therefore, be initiated based on the \emph{relative} weights of the generated chains. Moreover, the size of the set can easily be kept constant during growth: the algorithm will attempt to terminate (prune) chains that have a much lower effective statistical weight than the other candidate chains in the set and replace those with clones (enrichments) of ones with the largest weight. The computational effort in each MC step is, therefore, fixed by the selected number of candidate chains in the set. The performance of this approach will be analyzed through the simulation of polymer brushes.

\section{Algorithm}
Every MC step involves randomly selecting and removing a polymer chain from the system. The algorithm attempts to update the configuration of this chain, and inserts it back into the system at the end of the step. The CBMC approach to the former is the generation of a `probe' configuration $c_\mathrm{n}$ with weight $W_\mathrm{n}$, as described in algorithm A below. Also, a weight $W_\mathrm{e}$ for the old configuration $c_\mathrm{e}$ is obtained, as algorithm B describes. The acceptance rate of $c_\mathrm{n}$ as the new configuration is determined by a Metropolis-form probability\cite{Metropolis1953} $p_\mathrm{accep} = \mathrm{min} (1, W_\mathrm{n}/W_\mathrm{e})$. Starting at monomer $\ell=1$, the growth of a probe configuration $c = (r_1,\dots,r_L)$ proceeds as follows.  
\begin{itemize}
\item[A1]{Construct a set $\{\zeta_1, \dots,\zeta_k\}$ consisting of $k$ trial positions to insert monomer $\ell$. These positions are selected with a relative probability
\begin{equation}
p(\zeta_i) = \nu_\ell \exp(-\beta u^\mathrm{bond}_{\ell}(\zeta_i)) \label{eq:pbond},
\end{equation}
where $\nu_\ell$ is a normalization constant, $\beta$ the usual thermodynamic beta, and $u_\ell^\mathrm{bond}(\zeta)$ the bonding energy for monomer $\ell$, which is defined w.r.t. the position and orientation of the previous monomer. Note that the first monomer that is inserted ($\ell=1$) has a vanishing bonding energy, although for grafted polymers $u^\mathrm{bond}_1(\zeta)$ is the bonding energy to the grafting surface. The Rosenbluth factor associated with the set of trial positions is $w_\ell = \sum_{j=1}^k \exp [-\beta u^\mathrm{ex}_\ell (\zeta_j)]$, where $u^\mathrm{ex}_\ell(\zeta)$ is the potential energy of monomer $\ell$ which takes account of interactions with other monomers or fields. If $w_\ell>0$ then a position $r_\ell$ is selected from the set with probability $p_j = \exp [-\beta u^\mathrm{ex}_\ell (\zeta_j)] / w_\ell$} and the monomer is added to the chain.
\item[A2]{If a monomer was added, i.e. $w_\ell > 0$, then step A1 is repeated until the chain is complete ($\ell=L$).}
\end{itemize}
The algorithm for re-tracing an existing configuration, starting from $\ell=1$, is related to A.
\begin{itemize}
\item[B1]{Choose $\zeta_1 = r_\ell$ and generate the remaining set $\{\zeta_2, \dots,\zeta_k\}$ of $k-1$ other trial positions for monomer $\ell$ using Equation~(\ref{eq:pbond}). The Rosenbluth factor for this monomer is also given by $w_\ell = \sum_{j=1}^k \exp [-\beta u^\mathrm{ex}_{\ell} (\zeta_j)]$. The monomer is added to the chain at position $r_\ell$. }
\item[B2]{Step B1 is repeated for every next monomer, until the last monomer has been weighed ($\ell=L$).}
\end{itemize}
Using either algorithm, the resulting weight and energy of a chain are calculated as $W = \prod_{\ell=1}^L w_\ell$ and $U = \sum_{\ell=1}^L \left(u^\mathrm{bond}_\ell(r_\ell) + u^\mathrm{ex}_\ell(r_\ell)\right)$, respectively. Zero weight results for terminated chains.\\

The proposed algorithm takes a different approach from CBMC through the simultaneous construction of $M-1$ probe configurations while also re-tracing the existing chain. This yields a total of $M$ candidate configurations. There are no interactions between the monomers of different candidate configurations and separate Verlet lists may be used for each of them(as well as one for the rest of the system). Starting from monomer $\ell=1$, the growth of this set proceeds by repeating the following steps $L$ times.
\begin{itemize}
\item[C1]{Monomer $\ell$ is added to the $M-1$ probe chains that have not been terminated(as defined below) by applying algorithm $A1$. Monomer $\ell$ of the reference configuration is weighed using algorithm $B1$. The (Rosenbluth) weight of each unfinished chain $W_\ell = \prod_{\ell'=1}^\ell w_{\ell'}$ and its effective weight $W^{*}_\ell = 2^\gamma W_\ell$ is calculated. Here, $\gamma = y_p - y_e$, with $y_p$ the number of prune attempts, and $y_e$ the number of times cloning (enrichment) occurred(see below).}

\item[C2]{Chains with $W^*_\ell=0$ are terminated. The average effective weight $\tilde W_\ell^*$ of the other chains is calculated. Chains with $W_\ell^* < \tilde W_\ell^* / 2$ are marked for pruning. Those are terminated with probability $1/2$ and have $\gamma$ raised by 1 otherwise. The retraced chain, however, cannot be terminated and has $\gamma$ raised by 1 if it is marked for pruning.}
\item[C3]{The chain with the largest effective chain weight $W^\mathrm{*max}_\ell$ is selected. If there is more than one chain with this effective weight then one of these is selected randomly\footnote{To avoid comparison of floating-point numbers for equality one may select randomly from chains with $W^*_\ell > (1-\epsilon) W^\mathrm{*max}_\ell$ instead. Here, $\epsilon$ is a small number(e.g. $10^{-4}$).}. A terminated chain is now replaced by a clone of this selected chain. Both chains obtain half of the effective weight $W_\ell^*$ since $\gamma$ is now lowered by 1. If the retraced chain is cloned then this clone proceeds growing as a probe chain. This step is repeated until all terminated chains have been replaced. }
\end{itemize}

Algorithm C produces a set of $M$ configurations $c_m$, $1\leq m\leq M$ and corresponding (effective) weights $W^*(m) \equiv W_L^*(m)$. One of these candidates is selected with a relative probability
\begin{equation}
P_\mathrm{accep}(m) =\frac{ W^*(m) }{ \sum_{m=1}^M W^*(m)}. \label{eq:paccep}
\end{equation} 
This configuration updates the old polymer configuration in the system. 

\section{Justification of method}
For convenience, a path $x$ will be defined as a chain configuration as well as the remaining trial positions that were generated by algorithm A or B, 
\begin{equation}
x = \left((r_1,\dots,r_{L^{'}}), (\zeta^{k-1}_1,\dots,\zeta^{k-1}_{L^{'}})\right),
\end{equation}
where $c(x) := (r_1,\dots,r_{L^{'}})$ is the chain configuration and $\zeta^{k-1}_\ell := \{\zeta_{\ell,2},\dots,\zeta_{\ell,k}\}$ are the other $k-1$ trial positions $\zeta_{\ell,j}$ for each monomer $\ell$. Terminated chains are characterized by $L^{'}<L$ and $W(x)=0$, while successful chains correspond to paths with $L^{'}=L$. Every $x$ has a unique weight $W(x)$.\\

Consider the probability of generating a path $x$ with either algorithm A or B. Algorithm A generates a set of $k$ trial positions $\{\zeta_{\ell,1},\dots,\zeta_{\ell,k}\}$ for every monomer $\ell$, and selects $r_\ell$ from this set with a probability $e^{-\beta u(r_\ell)}/w_\ell(x)$. Algorithm B, on the other hand, fixes $\zeta_{\ell,1}=r_\ell$ and only needs to generate the $k-1$ remaining trial positions $\{\zeta_{\ell,2},\dots,\zeta_{\ell,k}\}$ associated with $x$. Whilst in algorithm A the position $r_\ell$ is selected with a probability $\exp [-\beta u^\mathrm{ex}_\ell (r_\ell)] / w_\ell$ , in algorithm B this probability iremarks s always 1. Therefore, the overall probability $p_\mathrm{A}(x)$ of generating a path $x$ with algorithm A is related to the probability $p_\mathrm{B}(x)$ of generating $x$ with algorithm B as
\begin{equation}
\frac{p_\mathrm{A}(x)}{p_\mathrm{B}(x)} = \prod_{\ell=1}^{L} \left(k \nu_{\ell} e^{-\beta u^\mathrm{bond}_\ell(r_\ell)}  \frac{e^{-\beta u^\mathrm{ex}_{\ell}(r_\ell)}}{w_\ell(x)}\right) = \nu \frac{e^{-\beta U(c(x))}}{W(x)}, \label{eq:papb}
\end{equation}
where Equation (\ref{eq:pbond}) was used and the $\nu \equiv \prod_\ell k \nu_\ell$ is a constant. Note that the probabilities of generating the unselected trial positions have cancelled each other in this equation.\\

\begin{figure}[ht] 
\centering\includegraphics[width = 8.5cm]{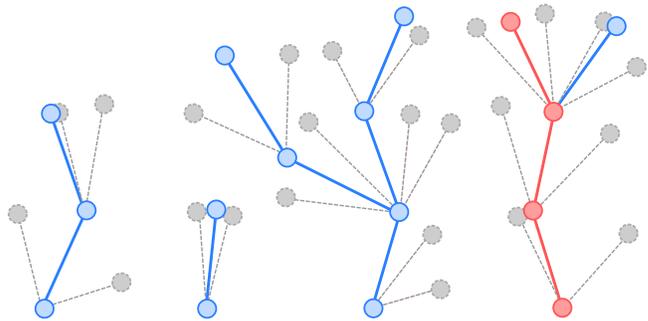}
\caption{Sketch of a set $s$ of paths explored in one Monte-Carlo step. The generated configurations are colored: probe configurations are blue and the retraced configuration is red. The number of trial position $k$ per monomer is set to three here, and the $k{-}1$ remaining directions are represented by the dashed grey lines. The number of growing chains $M$ is four at any time during the generation of the set. }  
\label{fig:RR}  
\end{figure}  

Algorithm C generates a set $s$ of paths during every MC step, as sketched in Figure \ref{fig:RR}. Despite the different coloring (red/blue) in the latter figure of the probe configurations and the retraced configuration, $s$ \emph{itself} does not hold information on which of the paths corresponds to the retraced configuration. The same set $s$ can result from any retraced configurations $c(x)$ as long as $x$ is a successful chain contained in $s$. Nevertheless, the probability $P_\mathrm{C}(s,x)$ of generating $s$ with algorithm C depends on the path $x$ to which the retracing algorithm B was applied. The rationale below will therefore be focused on comparing $P_\mathrm{C}(s,x)$ between different choices of $x$. For convenience, one may first relate $P_\mathrm{C}(s,x)$ to the probability $P_\mathrm{C*}(s)$ of generating $s$ with an (hypothetical) algorithm C* that does not retrace any existing configuration but applies algorithm A to all the paths. Consider selecting one of the $M$ (initially empty) paths and following its generation, starting from $\ell=1$. Given that $s$ is generated, the probability that \emph{this particular path} becomes $x$ is $1/(M \cdot  2^{y_e(x,s)})$, with $y_e(x,s)$ the number of times $x$ was cloned during its growth. The latter takes account for the fact that each cloning event branches off a new path that has an equal probability of becoming $x$. This defines $P_\mathrm{C*}(s,x) = P_\mathrm{C*}(s) /(M \cdot  2^{y_e(x,s)})$, which is the probability of generating $s$ while at the same time the selected path yields $x$.\\

In comparison with algorithm C*, algorithm C indeed selects one the $M$ initial paths and follows(directs) the growth of this path through the application of algorithm B. The probability that this path results in $x$ therefore increases by an additional factor $(p_\mathrm{B}(x)/p_\mathrm{A}(x)) \cdot 2^{y_p(x,s)}$. Here, $y_p(x,s)$ is the number of times $x$ was marked for pruning and accounts for the fact that a retraced chain cannot be terminated. Thus, the probability $P_\mathrm{C}(s,x)$ of generating $s$ while having $x\in s$ generated by retracing can be expressed as $P_\mathrm{C}(s,x) = P_\mathrm{C*}(s,x) (p_\mathrm{B}(x)/p_\mathrm{A}(x)) \cdot 2^{y_p(x,s)}$, which yields
\begin{eqnarray}
P_\mathrm{C}(s,x) =  P_\mathrm{C*}(s) \left[\frac{1} {M~2^{\gamma(x,s)}} \frac{p_\mathrm{B}(x)}{p_\mathrm{A}(x)} \right],\label{eq:Prws}
\end{eqnarray} 
recalling that $\gamma(x,s) = y_p(x,s) - y_e(x,s)$. Any move from an old polymer configuration $c_{a}$ to a new configuration $c_{b}$ must proceed through the generation of a set $s$ that contains $c_{a}$ as well as $c_{b}$, i.e. $x_{a},x_{b} \in s$, where $c(x_{a}) = c_{a}$ and $c(x_{b}) = c_{b}$. By defining $P^s_{c_{a} \rightarrow c_{b}}$ as the probability of moving from $c_{a}$ to $c_{b}$ \emph{via} the generation of $s$, one finds
\begin{eqnarray}
\frac{P^s_{c_{a} \rightarrow c_{b}}}{P^s_{c_{b} \rightarrow c_{a}}} = \frac{P_\mathrm{C}(s,x_{a}) \cdot P_\mathrm{accep}(x_{b},s)}{P_\mathrm{C}(s,x_{b}) \cdot P_\mathrm{accep}(x_{a},s)}, \label{eq:probflowa} 
\end{eqnarray}
where $P_\mathrm{accep}(x,s)$ is the probability of updating the polymer configuration to $c(x)$, given the set $s$. By combining Eqs.~(\ref{eq:Prws}) and (\ref{eq:papb}) it is possible to rewrite Equation (\ref{eq:probflowa}) as
\begin{equation}
\frac{P^s_{c_{a} \rightarrow c_{b}}}{P^s_{c_{b} \rightarrow c_{a}}} = e ^ {\beta (U(c_{a}) - U(c_{b}))} \frac{W(x_{a}) 2^{\gamma(x_{a},s)}}{W(x_{b}) 2^{\gamma(x_{b},s)}} \frac{P_\mathrm{accep}(x_{b},s)}{P_\mathrm{accep}(x_{a},s)}.\label{eq:probflowb} 
\end{equation}
and by using the selection criterion for a candidate configuration \ref{eq:paccep} the existence of (a variation of) superdetailed balance\cite{Frenkel1999a} is confirmed. Detailed balance is, therefore, satisfied globally as the total transition rate between $c_a$ to $c_b$ follows from all sets $s$ that obey this balance. 

\section{simulations}
The algorithm that is introduced here, which will be referred to as `swarm' confrontational-bias Monte-Carlo (SCBMC), is tested on a system of polymer brushes. Polymer brushes can be used as lubricants, adhesives, or to stabilize colloidal suspensions and, consequently, have been extensively studied by simulations and theory\cite{Murat1989,Milner1991,Lai1991,Netz1998,Patra2006,Binder2012,Loverso2013}. They form an interesting model system to test the algorithm since the simulation of dense, long brushes is computationally demanding. For the sake of generality, a simple freely-jointed chain model is considered here. The simulation box has horizontal dimensions $H\times H$ and is defined with periodic boundary conditions in this plane. Monomers are restricted in the vertical direction to $z\geq0$, such that $z=0$ defines the grafting surface to which $N=60$ polymers are grafted. Each chain is composed of $L$ hard spheres with diameter $d=1$, yielding a (dimensionless) grafting density of $\sigma = N d^2 / H^2$\cite{Murat1989}. For the initial configuration a random grafting of fully stretched chains is used.\\

\begin{figure}[ht] 
\centering\includegraphics[width = 8.5cm]{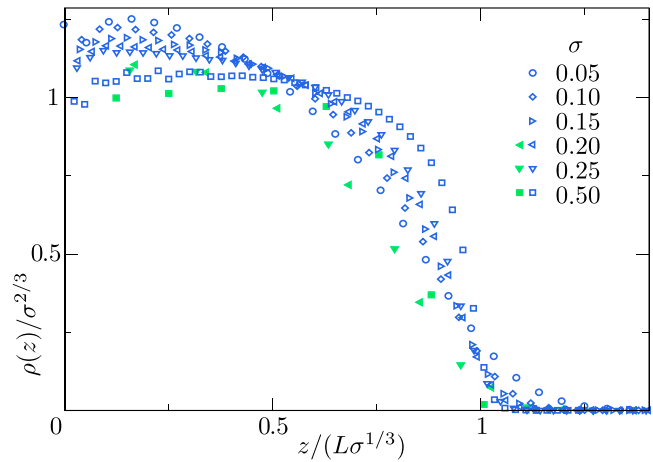}
\caption{Scaled monomer density plotted as a function of the scaled distance from the grafting plane. The blue(open) symbols denote data for $L=50$ and the green (closed) symbols correspond to $L=10$. Simulations ran for $10^4$ seconds for $L=50$ and $10^3$ seconds for $L=10$. Grafting densities $\sigma \geq 0.25$ ran 10 times longer.}  
\label{fig:scaling}  
\end{figure}  

All simulations were run on a single core of a Intel I7-6700 CPU in a desktop computer. Calculated density profiles have been checked for numerical accuracy by comparing to earlier work on similar systems\cite{Murat1989}. The data in Figure~\ref{fig:scaling}, furthermore, confirms the Alexander scaling of the obtained density profiles with the grafting density\cite{Alexander1977} for a range in $\sigma$ and two different polymer lengths. Deviations from a parabolic profile\cite{Milner1988,Netz1998} can be observed at extremely large grafting densities such as $\sigma=0.5$ when the system adopts a more rectangular profile. The latter is consistent with results from lattice simulations\cite{Coluzza2008}.\\

\begin{figure}[ht] 
\centering\includegraphics[width = 8.5cm]{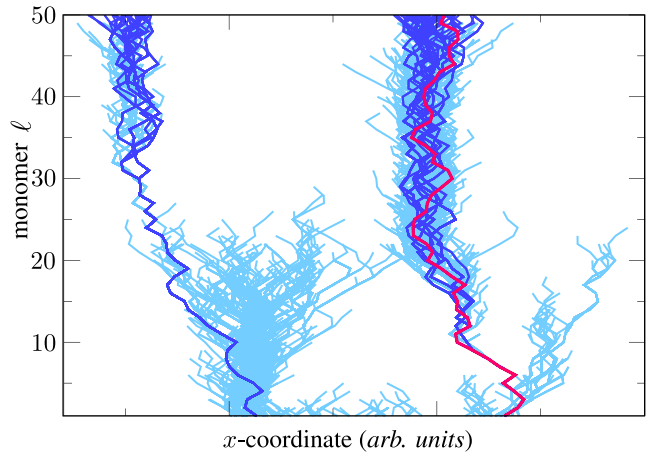}    
\caption{Sample chosen from simulations to demonstrate the pruning and cloning process applied on the swarm, showing the $x$-coordinate for the set of candidate chains that is generated in a MC step. The red curve corresponds to the retraced chain and the other completed chain configurations are darker blue. The lighter curves show the terminated chains. This figure is obtained from a simulation with $L=50$ and $\sigma=0.2$. The number of candidate chains $M$ was set to $150$.}  
\label{fig:RRswarm}  
\end{figure}  

Figure~\ref{fig:RRswarm} corresponds to a set of candidate configurations that the SCBMC algorithm generates during one MC step by plotting the $x$-coordinate of each monomer along the chains. The red curve shows the monomer positions of the retraced chain, while the other candidate configurations are drawn in darker blue. Although $M$ was set to $150$ in this simulation, the number of chain-termination events during the chain growth was much larger ($\approx 1800$  for the set here). Those configurations are drawn in lighter blue. Figure~\ref{fig:RRswarm}, therefore, demonstrates how a large number of $M$ enables the algorithm to explore `dead-end', as every terminated chain can easily be replaced by a clone of a more successful chain. \\

\begin{table}[ht]
\resizebox{\columnwidth}{!}{%
\begin{tabular}{c c|c|c|c|c|}
$$&$$&CBMC&$M=10$&$M=100$&$M=1000$\\ 
\hline
$L{=}10$,&$\sigma{=}0.20$&$19.972^{\pm0.008}$&$19.972^{\pm0.010}$&$19.94^{\pm0.03}$&$19.90^{\pm0.12}$\\
$L{=}50$,&$\sigma{=}0.05$&$199.97^{\pm0.14}$&$199.63^{\pm0.09}$&$199.73^{\pm0.14}$&$201.0^{\pm0.8}$\\
$L{=}50$,&$\sigma{=}0.20$&$602^{\pm 5} $&$550^{\pm 1}$&$552.1^{\pm0.7}$&$552^{\pm3}$\\
\hline
\end{tabular}}
\caption{The effect of varying the number of candidate chains $M$ on the calculated mean squared end-to-end distance $\langle R_z^2 \rangle$ and its standard error of the mean, $\sigma_m$. The number of trial directions per monomer, $k$, was set to $10$ here.}
\end{table}
\begin{figure}[ht]
\centering\includegraphics[width = 8.5cm]{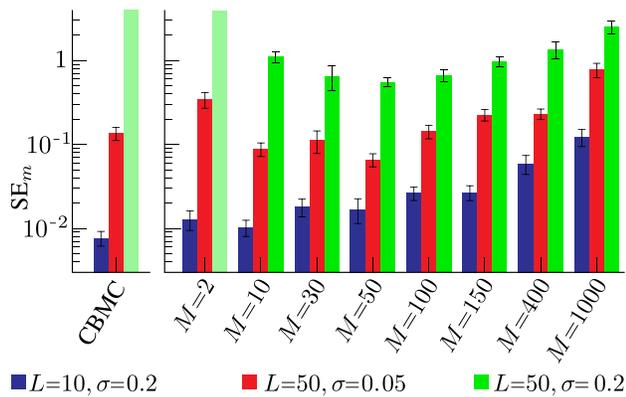}    
\caption{Comparison of the standard error of the mean, SE$_m$ between the CBMC method and the SCBMC method for an extended number of candidate chains $M$ w.r.t. to Table 1. All indicated confidence levels in SE$_m$ are estimated by bootstrapping.}
\label{fig:barplot}  
\end{figure}

The efficiency of the method in exploring phase space is analyzed by running approximately $10^4$-second simulations during which the mean-squared height of the end of the grafted chains $\langle R_z^2 \rangle$ was measured over 10 equal time intervals. The first of those is discarded for equilibration. For the shorter chain ($L=10$) the simulations only ran for approximately $10^3$ seconds in total. Table 1 shows results for $3$ different combinations of $L$ and $\sigma$. Most runs converged on the value of $\langle R_z^2 \rangle$, but a discrepancy is observed for the CBMC simulations of the longest and densest chains. An inspection of the final configuration revealed that the CBMC simulation did not completely move away from the initially stretched state within the given time. This rendered the value of  $\langle R_z^2 \rangle$ too large and underestimated the standard error due to correlations between samples. This also occurred for SCBMC simulations using $M=2$ but was absent for other tested values of $M$.\\

To get a better idea of the relative efficiency of different runs, Figure~\ref{fig:barplot} plots the error of the mean in $\langle R_z^2 \rangle$ for a wider range in $M$ than Table 1. For longer chains the results show that more candidates generally leads to better statistics, with an optimum value of $M\approx50$. This is most pronounced for $\sigma=0.2$. It can also be observed that simulations of short chains do not benefit from SCBMC here, which may be explained by the fact that within CBMC the success rate of updating configurations is already high for $L=10$.\\
\begin{figure}[ht] 
\centering\includegraphics[width = 8.5cm]{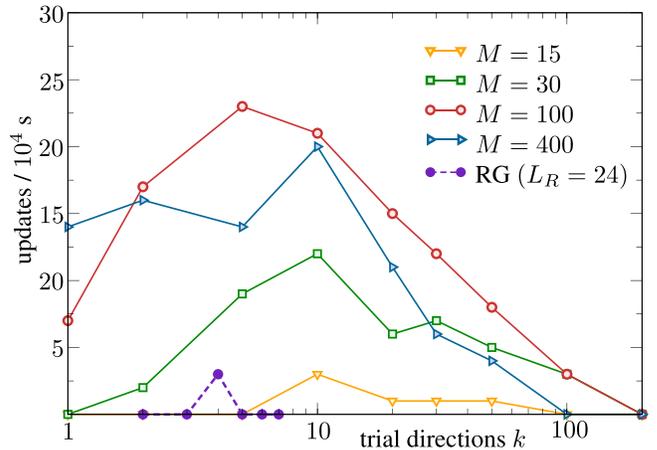}
\caption{Full chain updates during the course of $10^4$ s simulation time. The simulated brush has length $L=50$ and grafting density $\sigma=0.2$. The curves result from the SCBMC algorithm for given values of $M$, while varying $k$ along the horizontal axis. The dashed curve shows data from the recoil-growth algorithm, using a recoil length $L_R=24$.}  
\label{fig:update}  
\end{figure}  

Based on Figure~\ref{fig:RRswarm} it is expected that monomers lower in the chain are updated least frequently and will determine the longest correlation times. Consequently, the time it takes to update \emph{every} monomer position in the system may be used as an indicator for the efficiency of the algorithm. Figure~\ref{fig:update} plots the number of full updates during the run time of simulations with various algorithm parameters $M$ and $k$, using $L=50$ and $\sigma=0.2$. These results indicate the existence of an optimal choice for the number of trial directions $k$ for every choice of $M$. The CBMC approach did not produce any updates for any value of $k$, a full update from the stretched initial configuration did even not occur for simulations that ran 20 times longer, and so it was not included in this plot. For the data shown, simulations using $M\approx100$ and $k\approx5$ seem to achieve maximum efficiency. These findings qualitatively agree with the findings based on Table 1 and Figure \ref{fig:barplot}. Surprisingly, decent efficiency is also found for $k=1$ with a large number of candidate chains ($M=400$).\\

The efficiency of the SCBMC method may be understood from the possibility of cloning chains if some of them encounter `dead-ends'. Effectively, this redirects these chains into a direction that could be more successful. Since the recoil-growth (RG) algorithm \cite{Consta1999,Consta1999a} is based on similar ideas it is interesting to compare the efficiencies of both methods. The RG method extends on the CBMC method by enabling the probe chain to retract up to a maximum number of $L_R$ monomers once it meets a dead-end. This increases the chance that a successful chain configuration is found. An implementation of this method was checked for consistency with other methods. Full chain updates were attempted with a probability $1/2$, and partial chain updates otherwise. Note that partial chain updates cannot trigger counts of full updates directly, yet they were observed to be essential for equilibration of the system in test runs.  Since full updates of all monomer positions did not occur frequently, simulations were ran for $5$ times longer for the RG method. The parameters $L_R$ and $k$ were then optimized for efficiency. No full update of all monomers occurred for any $k$ except for $k=4$, for which optimal performance was observed by choosing $L_R\approx24$. The SCBMC method, therefore, seems to be less sensitive to the choice of various simulation parameters than the RG method. Although RG performs well compared to CBMC, the data in Figure~\ref{fig:update} also indicates that SCBMC outperforms RG by almost an order of magnitude in efficiency.

\section{Waste Recycling}
\begin{figure}[ht] 
\centering\includegraphics[width = 8.5cm]{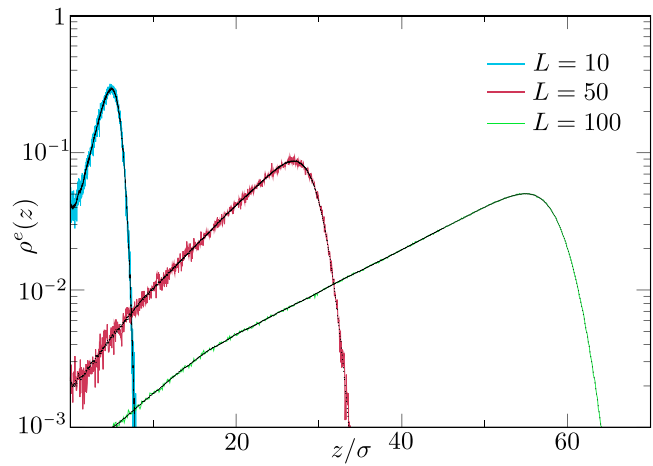}
\caption{Probability $\rho^e(z)$ of finding the end monomer at elevation $z$ from the grafting surface. While the colored curves are obtained without waste-recycling Monte Carlo (WRMC), the curves consisting of (small) black dots result from Equation~(\ref{eq:wasterecycling}). Simulations were performed for $\sigma=0.2$, using $M=150$, and $k=10$. For short ($L=10$), longer ($L=50$), and longest chains ($L=100$) these ran for $10^3$, $10^4$, and $10^6$ seconds respectively.}  
\label{fig:wrmc}  
\end{figure}  
Although the generation of a large number of candidate configurations is essential for optimal performance of the SCBMC algorithm, only one of these is recorded into the Markov chain of the system as the simulation proceeds. Recognizing the value of the discarded configurations is the central idea in waste-recycling Monte Carlo (WRMC), which re-uses these configurations for calculations of Boltzmann-weighted averages of observables \cite{Frenkel2004, Frenkel2006}. Because of detailed balance, one may record observables $A_n$ in succession to all Monte-Carlo steps $s_n$ that attempt to update polymer $n$, i.e.
\begin{equation}
\langle A_n \rangle \leftarrow \frac{\sum_{s_n} A_n}{ \sum_{s_n}}. \label{eq:statav}
\end{equation}
It was shown in Ref. \cite{Frenkel2006} that improved statistics can be obtained if a weighted average over the accepted as well as the rejected configurations is calculated. The relative weights herein are set by the acceptance probabilities of the candidate configurations, so equation~(\ref{eq:statav}) can be rewritten as
\begin{equation}
\langle A_n \rangle \leftarrow \left(\frac{1}{ \sum_{s_n}}\right) \sum_{s_n} \frac{\sum_{m=1}^M W^*(m) A_{n,m}}{  \sum_{m=1}^M W^*(m)}, \label{eq:wasterecycling}
\end{equation}
where $W^*(m)$ and $A_{n,m}$ are the effective weights and the value of observable $A_n$, respectively, corresponding to candidate configuration $m$. Equation~(\ref{eq:wasterecycling}) is used to sample the probability  $\langle \rho^e(z) \rangle= \sum_n \langle \rho^e_n(z) \rangle / N $ of finding the free end of a chain at a height $z$ from the grafting plane. Figure \ref{fig:wrmc} plots results for brushes with $\sigma=0.2$ for three different polymer lengths $L$. It can be observed that WRMC yields a strong reduction of the statistical noise in the determined profiles. The effectiveness of this combined approach is best illustrated by the acquired level of detail for the case $L=100$,  which is quite extraordinary for brushes with this density and length. WRMC may be particularly useful for determining the properties of the terminal monomer, as most of the generated candidate chains possess a unique configuration of this particular bead in the chain. On the other hand, properties related to monomers earlier in the chain may benefit less as their position is shared between multiple candidate chains. Nevertheless, waste recycling is basically `free' of computational effort to implement, so there is good reason to use it in combination with the introduced method.\\

\section{conclusion}
With respect to the physics that is discussed the results in this work agree with theoretical predictions for long and dense polymer brushes. More interesting is the efficiency of the introduced algorithm in simulating such systems. The simultaneous generation of a large set (swarm) of candidate configurations in each step of the dynamic MC algorithm can lead to an efficient sampling of phase space. An essential aspect of the introduced algorithm is the rigorous application of pruning and cloning(enrichment) during the generation of the set of candidate chains, which are balanced to keep the size of the swarm constant. It also ensures that all remaining configurations, including the reference configuration, have a comparable probability of being accepted. The optimal swarm size depends on the complexity of the system. This algorithm obtains accurate statistics from systems that do not seem to thermalize with the CBMC approach, and a strong increase of efficiency with respect to the recoil-growth method is observed. The parallelized chain generation can be distributed over multiple processing elements in future work\cite{Esselink1995}. Moreover, this method is tailored for waste-recycling Monte Carlo techniques, which enables the calculation of statistical averages from all generated chain configurations in the set.  The method is not limited to polymer systems only, and may be extended beyond chains to groups of particles in general. For CBMC this has already included mixtures of large and small particles \cite{Bolhuis1994,Biben1996}, phase-equilibrium calculations\cite{Dijkstra1994}, transition path sampling\cite{Dellago1998} and the structure of water\cite{Shelley1995} or ionic solutions \cite{Shelley1994}.
 
\acknowledgements
I would like to thank Daan Frenkel for his suggestion to consider incorporating waste-recycling Monte Carlo into the method. I gratefully acknowledge Alex Cumberworth for providing numerous valuable remarks on multiple versions of the manuscript.  
\end{document}